\documentstyle[a4,12pt]{article}

\newcommand{\bee}{\begin{equation}}
\newcommand{\ene}{\end{equation}}

\begin{document}
\title{On the Quantum Origin of the Mixmaster Chaos  
Covariance }
\maketitle
\author{{\bf Giovanni Imponente}$~^{1,2}$  and
{\bf Giovanni Montani}$~^{1,3}$ }\\
$~^1$ICRA---International Center for Relativistic Astrophysics, Piazza A.Moro 5 00185 Rome, Italy \\  
$~^2$Universit\`a degli Studi di Napoli ``Federico II'' and INFN-- Sezione di Napoli \\
$~^3$Physics Department, University of Rome ``La Sapienza''

\thanks{Authors can be contacted at e-mail addresses: imponente@icra.it, montani@icra.it.}

\begin{abstract}

Our analysis shows how the covariant chaotic behavior 
characterizing the evolution 
of the Mixmaster cosmology near the initial singularity 
can be taken as the semiclassical limit in the canonical quantization 
performed by the corresponding Hamiltonian representation. \\

\end{abstract}


\section{Cosmological Framework}

Though the Standard Cosmological Model (SCM) is based on the 
highly symmetric Friedmann-Lemaitre-Robertson-Walker (FLRW) solution 
of the Einstein equations, nevertheless
don't exist neither theoretically neither experimentally 
any evidences to prevent that
our universe underwent a more general dynamics in the very early stages 
of its evolution, and only in a later phase it isotropized reaching a complete
agreement with actual experimental data.\\ 
The simplest generalization of the FLRW dynamics consists
of the so-called Bianchi models, whose 
anisotropic evolution, for some of them, can be represented 
as a FLRW model plus a gravitational waves 
packet \cite{L74,GDY75}. 
Among this classification, the 
types VIII\footnote{All the considerations we will 
develop for the type IX apply also to the VIII one since, 
close to the singularity, they have the same morphology.} 
and IX appear as the most general ones allowed 
by the homogeneity constraint\footnote{IX's geometry is 
invariant under the $SO(3)$ group.} and  
their asymptotic 
evolution toward the cosmological singularity 
manifests a chaotic-like behavior \cite{BKL70}.
The cosmological interest lies in the IX model 
(the so-called Mixmaster \cite{M69}),  
which has the same space symmetries as the closed FLRW universe, 
and whose dynamics allows the line element 
to be decomposed as
\begin{equation}
\label{ds}
ds^2={ds_0}^2 -\delta_{(a)(b)}G^{(a)(b)}_{ik}dx^i dx^k
\end{equation}
where $ds_0$ denotes the line element of an isotropic universe having 
constant positive curvature, $G^{(a)(b)}_{ik}$ is a 
set of spatial tensors\footnote{These tensors satisfy the equations 
\[
G^{(a)(b);l}_{~ik~~;l}= -(n^2-3)G^{(a)(b)}_{~ik}, \quad 
G^{(a)(b)k}_{~~;k}=0, \quad
G^{(a)(b)i}_{~~i}=0 \, , 
\]
in which the Laplacian is referred to the 
geometry of the sphere of unit radius.}
and $\delta_{(a)(b)}(t)$ are amplitude functions, 
resulting small sufficiently far
from the singularity. \\
We dedicate 
our analysis to find a precise relation between the Mixmaster
deterministic chaos and the quantum 
behavior characterizing the Planckian era, showing 
how the invariant measure for the former, provided general Misner-Chitr\'e-like 
coordinates (MCl) \cite{KM97,M00}, is independent of the time gauge \cite{IM01}
and coincides with the stationary probability for the 
semiclassical limit of the latter.

\section{Billiard Representation}\label{hamiltonian}

Using generic MCl 
variables $(\xi, \theta, \tau)$ \cite{IM01},
the dynamics is described by the two-dimensional canonical variational 
principle
\begin{equation} 
\delta \int \left(   p_{\xi} {\xi}^{\prime} 
+  p_{\theta} {\theta}^{\prime} 
- {f}^{\prime}{\cal H}_{ADM} \right) d\eta =0\, ,
\label{q} 
\end{equation}
where $f$ is a generic function, 
\begin{equation} 
\label{n2}
{\cal H}_{ADM} = \sqrt{\varepsilon ^2 +U} ,\quad
\varepsilon ^2 = q^2{p_{\xi}}^2 +\frac{{p_{\theta}}^2}{q^2} 
\end{equation}
where $U(\xi, \theta,\eta)$ denotes the corresponding potential term 
and $q=q(\xi)\equiv \sqrt{\xi^2 -1}$.
Moreover the equation for the temporal gauge reads
\begin{equation} 
N\left(\eta\right)= \frac{12 D}{{\cal H}_{ADM}} e^{2f} \frac{df}{d\tau} {\tau}^{\prime} \, 
\label{rs} 
\end{equation} 
so that our analysis remains fully independent of the choice of the time 
variable until the form of $f$ and ${\tau}^{\prime}$ is not fixed.

For the following developments it is of 
key interest the following relation
\begin{equation} 
\label{t}
\frac{d\left({\cal H}_{ADM}f^{\prime}\right)}{d\eta} = \frac{\partial \left({\cal H}_{ADM}f^{\prime}\right)}{\partial\eta} \, .
\end{equation} 
The function $f\left(\eta\right)$ 
plays the role of a parametric function of time and actually the anisotropy 
parameters \cite{KM97}\cite{IM01} $H_i$ $(i=1,2,3)$ 
are functions of the variables $(\xi, \theta)$ only \cite{IM01}. 
In the domain $\Gamma_H$ where all the $H_i$ are simultaneously greater than 0, 
the potential term $U$ can be modeled by the 
potential walls\footnote{$\Theta _\infty \left(x\right) = \biggl\{ 
\begin{array}{lll} 
+ \infty & if & x < 0 \\ 
\quad 0 & if & x >  0 
\end{array} $}
\begin{equation}
\label{aa}
U_\infty = \Theta _\infty \left(H_1\right) + 
\Theta _\infty \left(H_2\right) + 
\Theta _\infty \left(H_3\right)   
\end{equation}
therefore, by (\ref{t}), in $\Gamma_H$ 
the ADM Hamiltonian 
becomes (asymptotically) an integral of motion
\begin{equation}
\label{bb}
\forall \{\xi, \theta\}\in{ \Gamma_H} 
\left\{ 
\begin{array}{lll} \displaystyle
\frac{\partial {\cal H}_{ADM}}{\partial f} &=& \displaystyle \frac{\partial E}{\partial f} =  0 \\
{\cal H}_{ADM}&=& \sqrt{\varepsilon ^2 + U} \cong \varepsilon =E 
\end{array}
\right .
\end{equation}
where $E$ is a constant.
In view of (\ref{bb}) 
the variational principle (\ref{q}) reduces to
\begin{equation}
\delta \int \left( p_{\xi} d\xi + p_{\theta} d\theta -Edf \right) 
=\delta \int \left(  p_{\xi} d\xi + p_{\theta} d\theta \right)=0 \, .
\label{cc}
\end{equation}
Following the standard Jacobi procedure to reduce this 
principle to a geodesic one we get,
for the closed domain region $\Gamma_H$, the Riemannian line element
\begin{equation}
\label{mm}
ds^2 =E^2 \left[ \frac{d{\xi }^2}{{\xi}^2 -1}+ \left(\xi^2 -1\right) d {\theta }^2 \right] .
\end{equation}
Since the above metric (\ref{mm}) has curvature scalar $R=-\frac{2}{E^2}$
the point-universe moves over a negatively curved bidimensional space on 
which the potential wall (\ref{aa}) cuts the region $\Gamma_{H}$; indeed
the invariant Lyapunov exponent for the dynamical 
flux associated to (\ref{mm}) reads 
$\lambda_v =1/E > 0$ \cite{IM01}.
The point-universe, bouncing on the potential walls,
is reflected from a geodesic to another one, making each of them unstable. 
By itself, the positivity of Lyapunov number is not enough to ensure 
the system chaoticity, since its derivation remains valid for any Bianchi 
type model, but for the Mixmaster case the potential walls reduce 
the configuration space to a compact region ($\Gamma_H$), 
ensuring a real chaotic behavior.

\section{Statistical Mechanics Approach}\label{statistical} 

For a Statistical Mechanics reformulation of the dynamics, 
we adopt in (\ref{q}) the restricted 
time gauge $\tau^{\prime}=1$, leading to the 
variational principle 
\begin{equation}
\label{px} 
\delta \int \left(   p_{\xi} \frac{d\xi }{df } +  p_{\theta} 
\frac{d\theta}{df } 
- {\cal H}_{ADM} \right) df = 0 .
\end{equation} 
In spite of this restriction, for any assigned 
time variable $\tau$ (i.e. $\eta$) there  
exists a corresponding function $f \left(\tau \right)$ 
(i.e. a set of MCl variables able to 
provide the scheme presented in Section \ref{hamiltonian}) 
defined by the (invertible) relation 
\begin{equation} 
\frac{df }{d\tau} = \frac{{\cal H}_{ADM}}{12 D }N\left(\tau \right) e^{-2f } . 
\label{qx} 
\end{equation} 
Hence the analysis to derive the invariant measure for the system follows 
the same lines presented in \cite{KM97,M00}. 
Indeed we got a suitable representation of the Mixmaster 
chaoticity in terms of a two-dimensional point-universe 
randomizing within $\Gamma_H$, admitting an ``energy-like'' 
constant of motion $\varepsilon = E$,  
then well-described by a {\it microcanonical ensemble}, whose 
Liouville invariant measure reads\footnote{$\delta\left(x\right)$ denotes the Dirac function.} 
\begin{equation} 
d\varrho = A \delta \left(E - \varepsilon \right)d\xi d\theta dp_{\xi }dp_{\theta }  
\, , \quad A=const. \, .
\label{ux} 
\end{equation} 
After the natural positions 
\begin{equation} 
\label{pos}
p_{\xi } = \frac{\varepsilon}{q}\cos\phi \, , \quad 
p_{\theta } = \varepsilon q\sin\phi  \, , 
\label{v} 
\end{equation} 
being $0\le\phi<2\pi$, 
and the integration over all values 
of $\varepsilon$\footnote{The dependence 
on the initial conditions doesn't contain any 
information about the system chaoticity.}, 
we arrive to the uniform invariant measure \cite{CB83,KM97}\cite{IM01a} 
\begin{equation} 
d\mu = w_{\infty}\left(\xi, \theta,\phi\right) d\xi d\theta d\phi 
\equiv \frac{1}{8\pi ^2}d\xi d\theta d\phi  \, . 
\label{x} 
\end{equation}

The key point of our analysis is that 
any stationary solution of the Liouville theorem, like (\ref{x}), remains valid 
for any choice of the time variable $\tau$; 
indeed in \cite{M00} the construction of the Liouville theorem with respect to the 
variables $(\xi, \theta,\phi)$ shows the existence of such properties 
even for the invariant measure (\ref{x}).

More precisely, in agreement with the analysis presented in \cite{M00}, 
during a free geodesic motion the asymptotic functions 
$\xi\left(f\right), \theta\left(f\right)$ 
and $\phi\left(f\right)$ are provided 
by the simple system
\begin{equation}
\label{eqmo1}
{\small
\frac{d\xi}{df}=q\cos\phi  , \quad
\frac{d\theta}{df}=\frac{\sin\phi}{q}  , \quad
\frac{d\phi}{df}=-\frac{\xi\sin\phi}{q}  
}
\end{equation}
and therefore over the reduced phase 
space\footnote{$S^1_{\phi}$ denotes the $\phi$-circle.} 
$\{\xi,\theta\}\otimes S^1_{\phi}$ the distribution $w_{\infty}$ 
behaves like the step-function 
\begin{equation}
\label{step}
w_{\infty}\left(\xi, \theta, \phi\right)=\left\{ \begin{array}{lll} \displaystyle
\frac{1}{8\pi^2} \, &\forall& \left\{ \xi, \theta, \phi\right\} \in \Gamma_H\otimes S^1_{\phi}  \\ 
0  \, &\forall& \left\{ \xi, \theta, \phi\right\} \not\in \Gamma_H\otimes S^1_{\phi} 
\end{array} 
\right. 
\end{equation}
stationary solution of the Liouville theorem 
\begin{equation}
\label{cont}
q\cos\phi \frac{\partial w_{\infty}}{\partial \xi}+
\frac{\sin\phi}{q}\frac{\partial w_{\infty}}{\partial \theta}-
\frac{\xi\sin\phi}{q}\frac{\partial w_{\infty}}{\partial \phi}=0 \, .
\end{equation}
If now we restrict our attention to the distribution function on the 
configuration space $\Gamma_H$
\begin{equation}
\label{conf}
\varrho\left(\xi, \theta\right)\equiv \int_0^{2\pi}w_{\infty}\left(\xi, \theta, \phi\right)d\phi \, ,
\end{equation}
by (\ref{cont}) we get for such reduced form the two dimensional 
continuity equation 
\begin{equation}
\label{cont2}
q\cos\phi \frac{\partial \varrho_{\infty}}{\partial \xi}+
\frac{\sin\phi}{q}\frac{\partial \varrho_{\infty}}{\partial \theta}=0 
\end{equation}
and the {\it microcanonical} solution on the whole configuration 
space $\{\xi, \theta\}$ then reads 
\begin{equation}
\label{steprho}
\varrho_{\infty}\left(\xi, \theta\right)=\left\{ \begin{array}{lll} \displaystyle
\frac{1}{4\pi} \qquad &\forall& \left\{ \xi, \theta\right\} \in \Gamma_H  \\ 
0  \qquad &\forall& \left\{ \xi, \theta\right\} \not\in \Gamma_H 
\end{array} 
\right. \, .
\end{equation}

\section{Quantum Origin of the Chaos}\label{quantum}

The main result of the above Section \ref{statistical} \cite{IM01,IM01a} 
is the proof that the chaoticity of the
Bianchi IX model above outlined is an intrinsic feature of its dynamics and 
not an effect induced by a particular class of references.
Appearing this intrinsic chaos close to the Big Bang,
we infer that it has strict relations with the indeterministic 
quantum dynamics the model performs in the {\it Planckian era}. 
The link between quantum and deterministic chaos is searched
in the sense of a semiclassical limit for the 
canonical quantization of the model.

Indeed the asymptotical principle (\ref{q}) can 
be quantized by a natural Schr\"oedinger approach \cite{KM97a} 
\begin{equation}
\label{sch}
i \hbar \frac{\partial \psi}{\partial \tau}=\hat{{\cal H}}_{ADM}\psi \, ,
\end{equation}
being $\psi=\psi(\tau,\xi,\theta)$ the wave function for the point-universe
and, implementing $\hat{{\cal H}}_{ADM}$ (see (\ref{n2})) to 
an operator\footnote{The non vanishing canonical commutation relations are
\[ \left[ \hat{\xi},\hat{p_{\xi}}\right] =i\hbar \, , \qquad \left[ \hat{\theta},\hat{p_{\theta}}\right] =i\hbar \, .
\]}, i.e.
\begin{eqnarray}
\label{op}
\xi &\rightarrow &\hat{\xi} \, , \quad \qquad  \quad \qquad  \theta \rightarrow \hat{\theta} \, ,  \nonumber \\
p_{\xi} &\rightarrow & \hat{p_{\xi}} \equiv -i \hbar \frac{\partial}{\partial \xi} \, , \quad 
p_{\theta} \rightarrow \hat{p_{\theta}} \equiv -i \hbar \frac{\partial}{\partial \theta} \, ,
\end{eqnarray} 
the equation (\ref{sch}) rewrites explicitly, in the asymptotic limit $U\rightarrow U_{\infty}$,
\begin{equation}
\label{sch1}
i \frac{\partial \psi}{\partial \tau} = \sqrt{\hat{\varepsilon}^2 +\frac{U_{\infty}}{\hbar^2}}~\psi  \, ,
\end{equation}
where we left $U_{\infty}$ to stress that the potential cannot be neglected 
on the entire configuration space $\{\xi, \theta\}$ and, being infinity out 
of $\Gamma_H$, it requires as boundary condition for $\psi$
to vanish outside the potential walls
$\psi\left(\partial \Gamma_H\right)=0  $.
Since the potential walls $U_{\infty}$ are time independent, 
a solution of (\ref{sch1}) can be taken in the form
\begin{equation}
\label{sol1}
\psi\left(\tau, \xi, \theta\right)= \sum_{n=1}^{\infty} c_n e^{-i E_n \tau / \hbar} \varphi_n\left(\xi, \theta\right)
\end{equation}
where $c_n$ are constant coefficients and we assumed a 
discrete ``energy'' spectrum because the quantum point-universe is restricted in the 
finite region $\Gamma_H$. Once taken an appropriate symmetric 
normal ordering prescription, the squared relation from (\ref{sch1}) 
and the position (\ref{sol1}) lead 
to the eigenvalue problem 
\begin{eqnarray}
\label{eig}\displaystyle
\left[-q\frac{\partial }{\partial\xi} q \frac{\partial }{\partial\xi}  
\displaystyle
-\frac{1}{q}\frac{\partial }{\partial\theta}\frac{1}{q}
\frac{\partial }{\partial\theta} \right] \varphi_n =  \qquad\qquad\qquad\nonumber\\
\displaystyle
= \frac{{E_n}^2-U_{\infty}}{\hbar^2} \varphi_n 
\equiv \frac{{E_{\infty}}^2_n}{\hbar^2} \varphi_n   . 
\end{eqnarray} 
The {\it quantum} equation (\ref{sch1}) is equivalent to the 
Wheeler-DeWitt one for the same Bianchi model, once separated the positive and negative
frequency solutions, 
with the advantage that now $\tau$ 
is a real time variable.

In what follows we search the semiclassical solution of this equation 
regarding the eigenvalue ${E_{\infty}}_n$ as a finite constant 
(i.e. we consider the potential walls as finite) and only 
at the end of the procedure we will take the 
limit for $U_{\infty}$ (\ref{aa}). \\
We infer that in the semiclassical limit when 
$\hbar \rightarrow 0$ and the 
{\it occupation number} $n$ tends to infinity 
(but $n\hbar$ approaches a finite value)
the wave function $\varphi_n\rightarrow \varphi$ 
and ${E_{\infty}}_n \rightarrow {E_{\infty}} $ 
so that we have
\begin{equation}
\label{expa}
\varphi\left(\xi, \theta \right) =\sqrt{r \left(\xi, \theta \right)} 
\exp\left\{ i\frac{S\left(\xi, \theta \right)}{\hbar}\right\} \, ,
\end{equation}
where $r$ and $S$ are functions to be determined. \\
Substituting (\ref{expa}) in (\ref{eig}) and separating the 
real from the complex part we get two independent equations, i.e.
\begin{equation}
\label{1eq}
{E_{\infty}}^2 = \underbrace{ q^2 \left( \frac{\partial S}{\partial \xi}\right)^2 + 
\frac{1}{q^2} \left( \frac{\partial S}{\partial \theta}\right)^2 }_{\rm classical~term} 
+\underbrace{~\ldots~}_{O(\hbar^2)} 
\end{equation}
\begin{equation}
\label{2eq}
0 = \underbrace{q \frac{\partial }{\partial \xi} \left( q~r \frac{\partial S}{\partial \xi} \right) 
+\frac{1}{q^2}\frac{\partial }{\partial \theta} \left( r \frac{\partial S}{\partial \theta}\right)}_{O(1/\hbar)}  \, .
\end{equation}
The dominant term in (\ref{1eq}) reduces to 
the Hamilton-Jacobi equation 
and its solution can be easily 
checked to be\footnote{The discontinuity of this 
function on the boundary of $\Gamma_H$ is due to the model 
adopted and doesn't affect the probability distribution.} 
\begin{equation}
\label{esse}
S\left(\xi, \theta\right) = \int\left\{ \frac{1}{q}\sqrt{E^2_{\infty}
 - \frac{k^2}{q^2}}~d\xi +k ~d\theta\right\} 
\end{equation} 
where $k$ is an integration constant.
Through the identifications   
\begin{equation}
\label{ide}
\frac{\partial S}{\partial \xi} =p_{\xi} \, , \, \frac{\partial S}{\partial \theta}=p_{\theta} 
\,\Longleftrightarrow \, S=\int\left( p_{\xi} d\xi + p_{\theta} d \theta \right) \, ,
\end{equation} 
(\ref{1eq}) is reduced to a mere algebraic constraint 
which is the asymptotic one 
${\cal H}_{ADM}^2=E^2\equiv E^2_{\infty}$ and is solved 
by (\ref{v}) replacing $\varepsilon=E_{\infty}$,
whose compatibility with (\ref{esse}) and (\ref{ide}) is then 
obtained using the equations of motion (\ref{eqmo1}) 
which provide
\begin{equation}
\label{eqmo}
\frac{d\xi}{d\phi}= - \frac{\xi^2}{\xi^2-1}{\rm ctg}\varphi \Rightarrow 
\sqrt{\xi^2-1}\sin\varphi =c \, , 
\end{equation}
where $c=const.$; the required compatibility 
comes from the identification $k=E_{\infty}c$.
Since $E_{\infty}\rightarrow i \infty$ outside $\Gamma_H$
the solution $\varphi\left(\xi, \theta\right)$ vanishes 
out of the billiard.\\
The substitution in (\ref{2eq}) of the positions 
(\ref{pos}) leads to the new equation
\begin{equation}
\label{cont3}
q\cos\phi \frac{\partial r}{\partial \xi}+
\frac{\sin\phi}{q}\frac{\partial r}{\partial \theta}=0 \, ,
\end{equation}
which coincides with (\ref{cont2}), 
provided the identification $r\equiv\varrho_{\infty}$; 
this result ensures the correspondence between the statistical and the 
semiclassical quantum analysis. \\

\end{document}